\documentclass[12pt]{spie}

\usepackage{amsmath,amsfonts,amssymb}
\usepackage{graphicx}
\usepackage[colorlinks=true, allcolors=blue]{hyperref}

\title{Progress in the Construction and Testing of the Tianlai Radio Interferometers}

\author[a,h]{Santanu Das} 
\author[a]{Christopher J. Anderson}
\author[b]{Reza Ansari}
\author[b]{Jean-Eric Campagne}
\author[b]{Daniel Charlet}
\author[c]{Xuelei Chen}
\author[e]{Zhiping Chen} 
\author[a]{Aleksander J. Cianciara}
\author[o]{Pierre Colom}
\author[c,d]{Yanping Cong}
\author[a]{Kevin G. Gayley}
\author[f]{Jingchao Geng} 
\author[l]{Jie Hao}
\author[c,d]{Qizhi Huang}
\author[a]{Celeste S. Keith} 
\author[c]{Chao Li}
\author[c,d]{Jixia Li} 
\author[i]{Yichao Li}
\author[f]{Chao Liu}
\author[e]{Tao Liu} 
\author[n]{Christophe Magneville}
\author[h]{John P. Marriner} 
\author[o]{Jean-Michel Martin}
\author[b]{Marc Moniez}
\author[a]{Trevor M. Oxholm}
\author[m]{Ue-Li Pen}
\author[b]{Olivier Perdereau}
\author[g]{Jeffrey B. Peterson} 
\author[c]{Huli Shi} 
\author[l]{Lin Shu}
\author[h]{Albert Stebbins} 
\author[c,d]{Shijie Sun} 
\author[a]{Peter T. Timbie}
\author[p]{Steve Torchinsky}
\author[j]{Gregory S. Tucker}
\author[f]{Guisong Wang}
\author[e]{Rongli Wang}
\author[m]{Xin Wang}
\author[c]{Yougang Wang}
\author[c]{Fengquan Wu} 
\author[c]{Yidong Xu} 
\author[c,d]{Kaifeng Yu}
\author[q]{Jiao Zhang}
\author[e]{Juyong Zhang} 
\author[k]{Le Zhang}
\author[e]{Jialu Zhu}
\author[c,d]{Shifan Zuo}

\affil[a]{Department of Physics, University of Wisconsin \textemdash{} Madison,
Madison, WI 53706, USA}

\affil[b]{LAL, Univ. Paris-Sud, CNRS/IN2P3, Universit\'e Paris-Saclay, Orsay, 
France}

\affil[c]{National Astronomical Observatories Chinese Academy of Sciences,
Beijing 100101, P. R. China}

\affil[d]{University of Chinese Academy of Sciences Beijing 100049, P. R. China}

\affil[e]{Hangzhou Dianzi University, Hangzhou, P. R. China} 

\affil[f]{The 54th Research Institute of the China Electronics Technology Group Corporation (CETC54),
Shijiazhuang, P. R. China}

\affil[g]{Department of Physics, Carnegie Mellon University,
Pittsburgh, PA 15213, USA} 

\affil[h]{Fermi National Accelerator Laboratory, Batavia, IL 60510, USA}

\affil[i]{University of KwaZulu-Natal (UKZN), Durban, S.A.}

\affil[j]{Department of Physics, Brown University, Providence, RI 02912, USA}

\affil[k]{Shanghai Jiao Tong University, Shanghai, P. R. China}

\affil[l]{Institute of Automation, Chinese Academy of Sciences, Beijing, P. R. China}

\affil[m]{Canadian Institute for Theoretical Astrophysics, Toronto, ON M5H 3H8, Canada}

\affil[n]{DSM/Irfu/DPhP, CEA-Saclay, F-91191 Gif-sur-Yvette Cedex, France}

\affil[o]{CNRS, Observatoire de Paris (LESIA, GEPI), Universit\'e PSL,
61 Ave de l'Observatoire, 75014 Paris, France}

\affil[p]{CNRS/IN2P3, Observatoire de Paris, Laboratoire Astroparticule \& Cosmologie (APC), Universit\'e Paris-Diderot, 75013 Paris, France}

\affil[q]{College of Physics and Electronic Engineering, Shanxi University, 92 Wucheng Road, Taiyuan, 030006 Shanxi, China}

\authorinfo{Further author information: (Send correspondence to Santanu Das.)\\Santanu Das: E-mail: sdas33@wisc.edu}

\begin{document}

\maketitle
\begin{abstract}
The Tianlai Pathfinder is designed to demonstrate the feasibility 
of using wide field of view radio interferometers to map the density
of neutral hydrogen in the Universe after the Epoch of Reionizaton.
This approach, called 21~cm intensity-mapping, promises an inexpensive
means for surveying the large-scale structure of the cosmos. The Tianlai Pathfinder presently
consists of an array of three, 15~m $\times$ 40~m cylinder telescopes and an
array of sixteen, 6~m diameter dish antennas located in a radio-quiet part
of western China. The two types of arrays were chosen to determine the advantages
and disadvantages of each approach.  The primary goal of the Pathfinder is to make 3D maps by surveying neutral hydrogen over
large areas of the sky 
in two different redshift ranges: first at $1.03 > z > 0.78$ ($700 -
800$~MHz) and later at $0.21 > z > 0.12$ ($1170 - 1270$~MHz).  The most significant challenge
to $21$~cm intensity-mapping is the removal of strong foreground radiation that
dwarfs the cosmological signal. It requires exquisite
knowledge of the instrumental response, i.e. calibration. In this
paper we provide an overview of the status of the Pathfinder and discuss 
the details of some of the analysis that we have carried out to 
measure the beam function of both arrays.  We compare electromagnetic simulations of the arrays
to measurements, discuss measurements of the gain and phase stability of the instrument, and provide a brief overview of the data processing pipeline.
\end{abstract}



\section{Introduction}
In the past few decades the cosmic microwave background (CMB) has been one of the 
primary observational tools  for cosmology. Experiments like {\sl WMAP}, {\sl Planck}, etc. measure the temperature fluctuations in the CMB extremely accurately.  All these experiments help us to constrain the standard model parameters \cite{Ade2016, Das2014, Das2015a}. However, the CMB gives only 2D 
information from the sky and CMB measurements are not very sensitive to the recent expansion history \cite{Das2013, Das2014b} which is most affected by accelerated expansion / dark energy.  Optical galaxy surveys have given us the most information on the dark energy phenomenon but it is expensive to extend densely sampled optical redshift surveys beyond a redshift of $z\sim 2$.
In the last decade there has been an explosion of interest in $21$~cm intensity-mapping observations since measurements of the $21$~cm line from neutral hydrogen can in principle be used to make 3D maps of matter in the universe at all redshifts up into the ``dark ages" ($z \lesssim 10^2$), even before galaxies have formed.  However the $21$~cm intensity-mapping technique has not yet been demonstrated in practice.

Neutral hydrogen, HI, consists of an electron bound to a proton. Both the electron and the proton have intrinsic magnetic dipole moments as a result of their spin; the spin-spin interaction results in a slightly higher energy of the hydrogen when the spins are parallel compared to the energy when the spins are anti-parallel. The fact that only parallel and anti-parallel spin states are allowed is a result of the quantum mechanical discretization of the total angular momentum of the system. The hydrogen atom emits or absorbs a photon at 21~cm when transitioning between these two states.  This transition is highly forbidden, with an extremely long mean lifetime for the excited state of around $10$ million years.  As a result, a significant signal from neutral hydrogen results only when there is a large amount of neutral hydrogen to observe, such as from astronomical sources.

The 21~cm line from neutral hydrogen is unique in cosmology because for $\lambda>21$~cm it is the dominant astronomical line emission.  With high confidence the wavelength of a spectral feature can be converted to the redshift without determining the source of the emission.  This method allows one to determine the redshift of ensembles of galaxies at a small range of redshifts (intensity-mapping) unlike traditional galaxy surveys which require the the determination of redshifts of individual galaxies containing heavy elements produced by star formation.  For this reason, HI intensity mapping can relatively inexpensively track inhomogeneities in the matter from the cosmic dark ages to the present time~\cite{Cianciara2017}. 



Neutral hydrogen intensity-mapping has been developed over the past decade as a possible means to measure large-scale structure in the Universe in a relatively inexpensive way \cite{Abdalla2005,Peterson2006, Morales2008, Chang2008,Mao2008}.   The traditional galaxy redshift survey is a time-consuming process that requires detecting a large number of individual galaxies and determining their positions and redshifts. Such a survey has much finer spatial resolution than is required for measuring the power spectrum in the linear regime of cosmic growth.  The nonlinear scale corresponds to a co-moving scale of $\sim 10$ Mpc, a scale that typically includes $\sim 100 - 1000$ galaxies.  Near redshift $1$, this scale subtends an angle of $\sim 20$ arcminute on the sky, transverse to the line of sight.  The fundamental idea behind 21~cm intensity-mapping is to measure the combined HI emission from many galaxies at once, simultaneously reducing the required angular resolution of the telescope and increasing the signal-to-noise ratio.  While the 21~cm signal has been exploited to conduct galaxy redshift surveys of individual galaxies out to $z\sim0.1$ \cite{Zwaan2001, Martin2010} beyond this redshift current single dish radio telescopes do not have sufficient angular resolution and sensitivity to make surveys of individual galaxies.

The most significant challenge to 21~cm intensity-mapping is extracting the HI signal from strong Galactic foregrounds that are $\sim 1000$ times greater.   In principle, the foregrounds should be separable from the signal because the spectra are very different: the foregrounds are dominated by synchrotron radiation and free-free emission, which have smooth, power-law spectra, while the HI signal from clumps of HI emitting at different redshifts forms a `spikey' spectrum.  The first measurements of the HI power spectrum using 21~cm intensity-mapping, reported beginning in 2010, were made with the Green Bank Telescope (GBT) at $z\sim 0.8$. HI maps were cross-correlated with maps of galaxy number counts from the DEEP2 and WiggleZ galaxy redshift surveys \cite{Chang2008, Chang2010, Masui2013, Switzer2013}.   As demonstrated at the GBT, frequency-dependent cross-coupling from strong polarized sources into the intensity response of the telescope can introduce frequency-dependent contamination into the signal. Minimizing and measuring these spurious effects is critical  \cite{Switzer2015}.

To survey large swaths of the sky with adequate signal-to-noise requires dedicated instrumentation. Both single dish and interferometric approaches are being developed.  BINGO\cite{Battye2013, Dickinson2014} will build an array of $\sim 50$ feed antennas for the focal plane of an off-axis reflector.  The 19-beam L-band focal plane for FAST could be used for an intensity-mapping survey as well\cite{Bigot2016}. A 7-beam array is under development for the $700 - 900$~MHz band of the GBT\cite{Chang2016}.  Although single-dish instruments may have less chromatic response than do interferometers, and hence have a significant advantage for the removal of foregrounds and instrumental effects, it has proved difficult to increase the mapping speed of single-dish instruments to compete with that of large-N interferometers.  Furthermore, the expense of building a large aperture single dish that can be dedicated to 21~cm intensity-mapping has proved prohibitive. As a result, most 21~cm intensity-mapping instruments are interferometers \cite{Ansari2008,Seo2010,Ansari2012,Xu2015,Bull2015}
and include cylindrical reflectors (Pittsburgh CRT\cite{Bandura2011},  CHIME\cite{Bandura2014}, the Tianlai cylinder array\cite{Chen2012}) as well as arrays of single dishes (Tianlai dish array and HIRAX\cite{Newburgh2016}). The  designs of these interferometers are related to those of wide-field arrays under development for studies of the Epoch of Reionization (EoR). EoR arrays operate at lower frequencies ($\sim 100$~MHz) to observe emission from HI at $z\sim 10$ and include LOFAR\cite{Yatawatta2013}, MWA\cite{Pober2016}, PAPER\cite{Ali2015}, and HERA\cite{Neben2016, Ewall2016}.

This paper describes the Tianlai Pathfinder, two low-frequency radio interferometers dedicated to 
21~cm intensity-mapping at redshifts $1.36 > z > 0$, corresponding to HI emission from 600~MHz to 1420~MHz \cite{Chen2012, Xu2015}. The paper is organized as follows. In Sec.~\ref{sec:Instrument}, we describe the design details of the Pathfinder. Some of the recent measurements and analysis are described in Sec.~\ref{sec:description}. 
Sec.~\ref{sec:tlpipe} is dedicated to providing a brief outline of the Tianlai software pipeline and finally Sec.~\ref{sec:conclusion} concludes the paper with a summary of future plans.

\section{The Tianlai Arrays}
\label{sec:Instrument}

\begin{figure}[t]
\includegraphics[width=0.49\textwidth,trim =  180 30 87 10, clip]{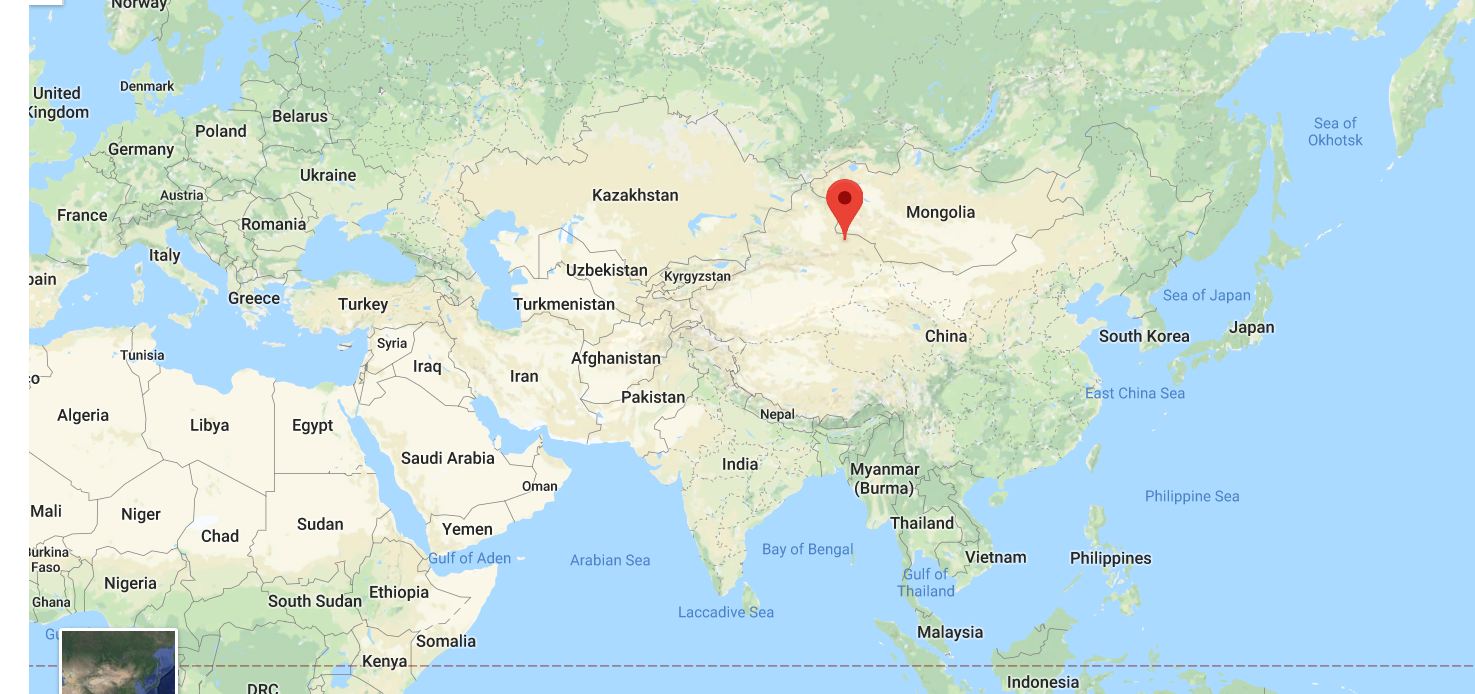}
\includegraphics[width=0.49\textwidth,trim =  0 10 0 10, clip]{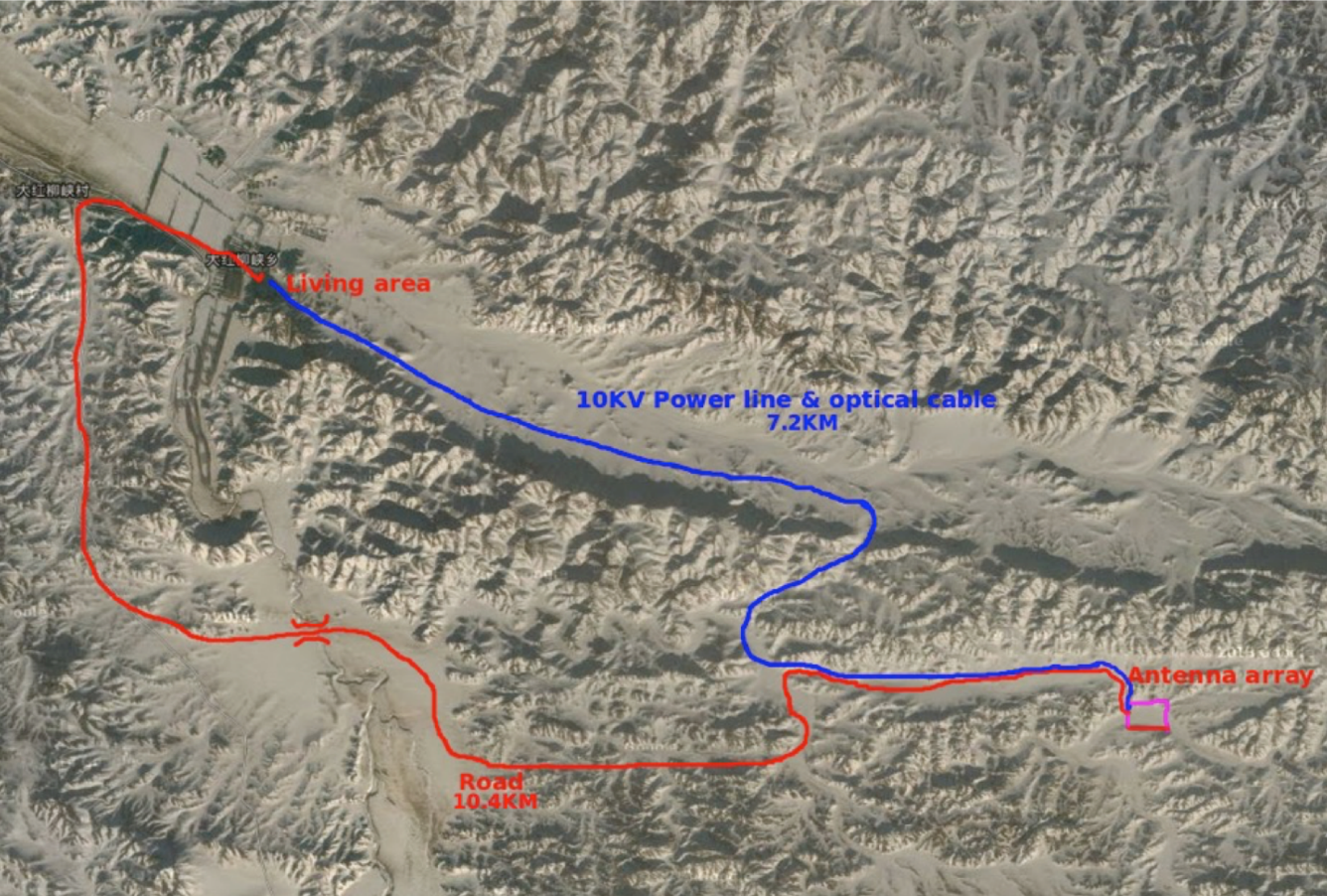}
\caption{\label{fig:location} Tianlai Pathfinder is located in Hongliuxia,
Balikun County, Xinjiang Autonomous Region in northwest China ($44^\circ 9' 9.66''$~N $91^\circ 48' 24.72''$~E).
Left : Location of the Tianlai Pathfinder. Right : Location of the Pathfinder and living area and the correlator building on a map. The blue line shows the power line and the optical cables connecting these two locations.  The red line represents the road used for transport.
}
\end{figure}

The Tianlai Pathfinder is located in Hongliuxia, Balikun County, Xinjiang Autonomous Region, in northwest China ($44^\circ 9' 9.66''$~N $91^\circ 48' 24.72''$~E), a radio-quiet site chosen after a survey of more than 100 candidate sites in China. Fig.~\ref{fig:location} shows the location 
of the array on the map. The name `Tianlai' means `heavenly sound' in Mandarin. 

\begin{figure}
\includegraphics[width=0.98\textwidth]{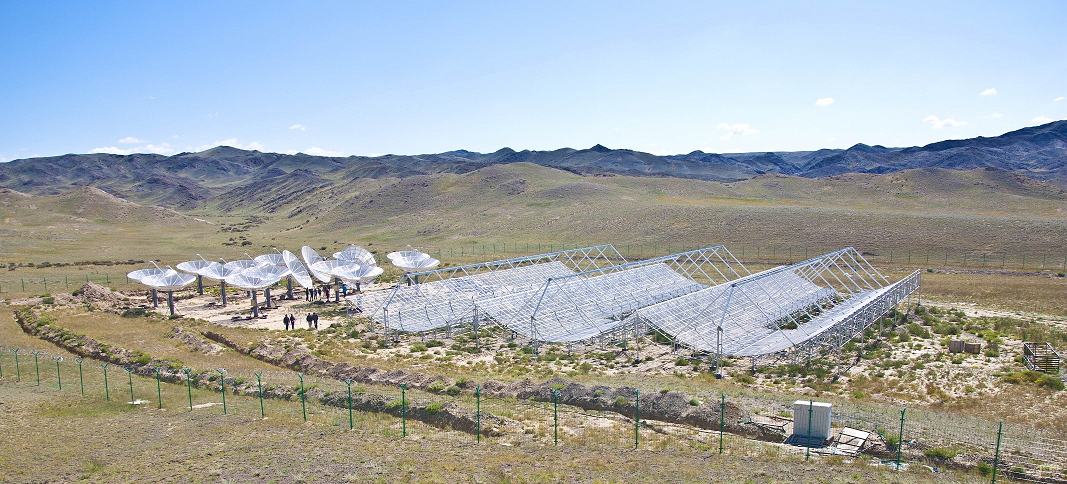}
\caption{\label{fig:arialview} The photograph shows the Tianlai Pathfinder dish and cylinder arrays.  The dish array is an array of 16, fully-steerable, on-axis dish antennas, each 6 m in diameter.  A dual-linear polarization feed is located at the focus of each dish. The cylinder array consists of 3, fixed, cylindrical reflectors, each 15~m wide and 40~m long, placed side-by-side (in the E-W direction) with their long axes oriented along the N-S direction. A linear array of dual-linear polarization feeds is arranged along their focal lines. The photograph is taken looking toward the southeast.}
\end{figure}

The Tianlai project is divided into three stages - Pathfinder,
Pathfinder+, and Full Array. Presently we are working on the
Pathfinder project. The objective is to implement large
interferometric arrays to obtain high fidelity 3D images of the northern sky. If successful, the project can be extended to the next stages.  The Pathfinder comprises two arrays, one consisting of dish antennas and the other of cylinder reflectors (see Fig.~\ref{fig:arialview}). Construction of the Pathfinder was completed in 2016 and it is now undergoing the commissioning process. 

For each array, the feed antennas, amplifiers, and reflectors are designed to operate from 600~MHz to 1420~MHz.  The instrument can be tuned to operate in an RF bandwidth of 100~MHz centered at any frequency in this range by adjusting the local oscillator frequency in the receivers and replacing the band pass filters.  Currently, the Pathfinder operates at $700 - 800$~MHz, corresponding to $1.03 > z > 0.78$. The system noise temperatures for both the dish and cylinder arrays are $\sim 80 - 85~{\rm K}$.


In order to avoid radio-frequency interference (RFI),
the correlator building and living quarters, were built about $10.4\,\text{km}$ away from the site (by road). The power line and optical fiber cables of about 7.2~km connect the correlator building with the antenna arrays. The power line and roadway connecting these two places are shown in Fig.~\ref{fig:location}. 

\subsection{Tianlai Dish Array}

The Tianlai dish array consists of $16$ on-axis dishes. Each has an aperture of $6\,\text{m}$. The array is roughly close-packed, with center-to-center spacings between neighboring dishes of about $8.8$~m. The dishes are equipped with dual, linear-polarization receivers. They are mounted on Alt-Azimuth mounts, and motors are used to control them electronically. The motors can steer the dishes to any direction in the sky above the horizon. The drivers are not specially designed for tracking celestial targets with high precision.  Instead, in the normal observation mode we point the dishes at a fixed direction and perform drift scan observations. The Alt-Azimuth drive provides flexibility during commissioning for testing and calibration. 

\begin{figure}
\includegraphics[width=0.4\textwidth,trim =  20 10 330 10, clip]{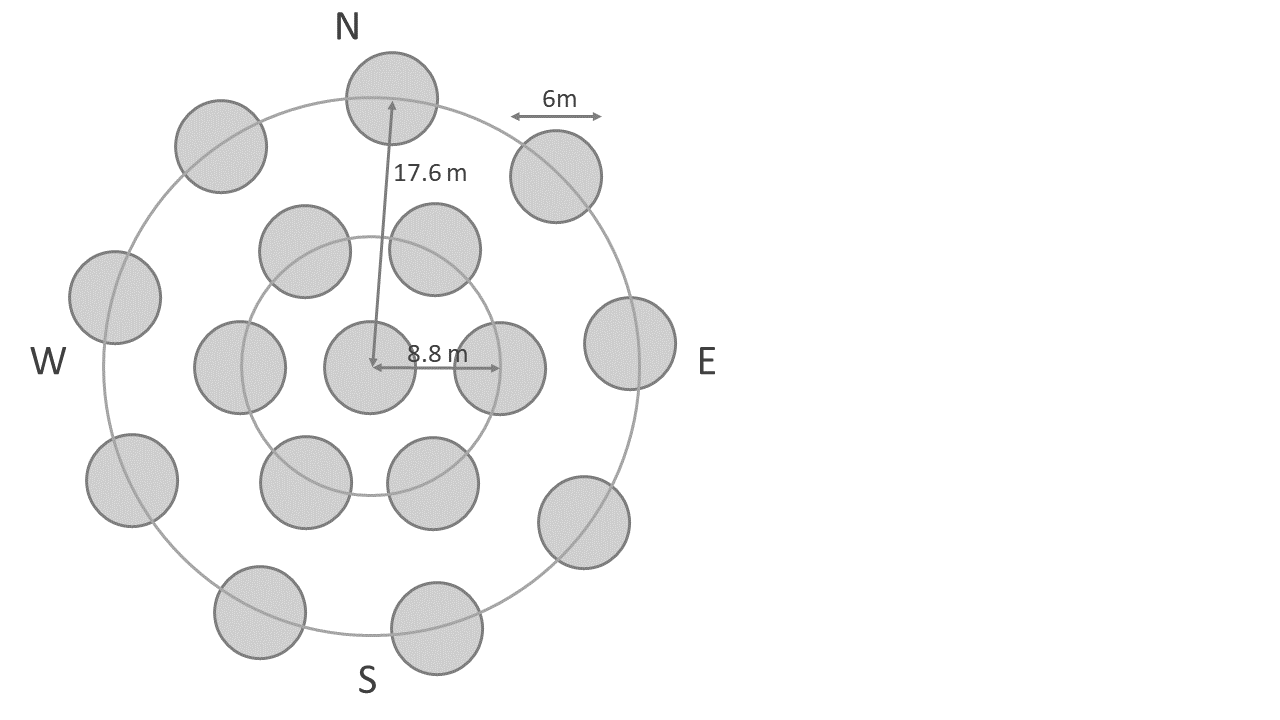}
\includegraphics[width=0.6\textwidth,trim =  100 40 220 10, clip]{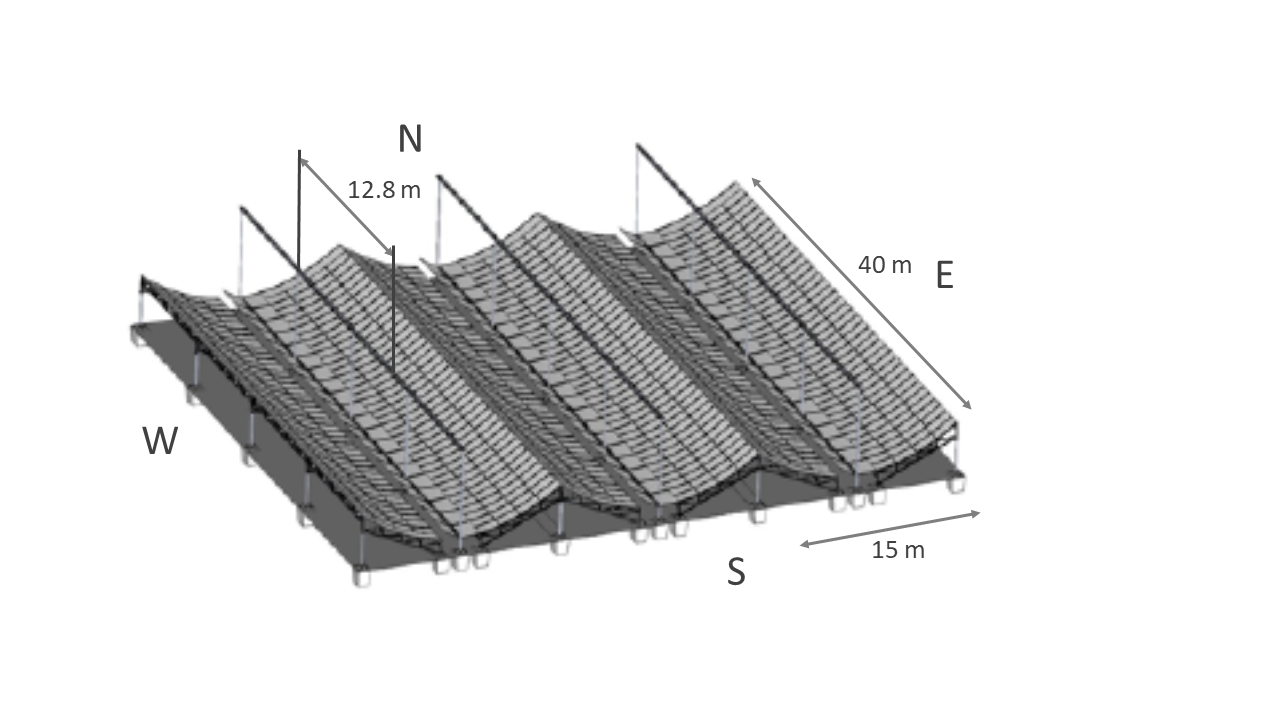}
\caption{\label{fig:scimatic}A schematic diagram of the Tianlai dish and cylinder arrays are shown here. The dishes are arranged in two concentric circles of radius $8.8\,$m and $17.6\,$m around a central dish. The cylinder array consists of 3 cylindrical reflectors. Currently, the central $12.8\,$m parts of the cylinders are equipped with receiver feeds. The number of feeds for the reflectors are 31, 32 and 33 respectively (from East to West).}
\end{figure}

We have adopted a circular configuration for the current Tianlai Pathfinder dish array (see Fig.~\ref{fig:scimatic}). One antenna is positioned at the center and the remaining $15$ antennas are arranged in two concentric circles around it. It is well known that the baselines of circular array configurations are quite independent and have wide coverage of the ($u$, $v$) plane. A comparison of the different configurations considered for the Tianlai dish array and the performance of the adopted configuration can be found in Zhang {\it et al.} (2016a) \cite{jzhang2016a}.  The Tianlai dishes are lightweight and the mounts are detachable, so, in future, we can move the dishes to new configurations if desired. 

The visibilities for the dish array (i.e. $32$ auto-correlations and $496$ cross-correlations) are computed by the data-acquisition (DAQ) system and saved on hard drives. 
The dishes currently observe the frequency band $700 - 800\,\text{MHz}$
($1.03>z>0.78$) in 512 frequency channels ($\delta \nu=244.14\,\text{kHz}$,
$\delta z=0.0002$).

\subsection{Tianlai Cylinder Array}

The Tianlai Pathfinder cylinder array has three adjacent cylindrical reflectors oriented in the North-South direction. Each of the cylinders is $15\,\text{m}$ wide and $40\,\text{m}$ long. At present the cylinders are equipped with a total of $96$ dual polarization feeds which do not cover the full length of the cylinders. The reflector centers are $15.2~$m apart.  The reflector shape focuses radiation in the  transverse direction, but not along the longitudinal direction, of each antenna.  In this way,  each feed antenna placed along the cylinders' focal line receives radiation from a fan-shaped  beam oriented along the meridian.  The three cylinders have 31, 32, and 33 feed antennas  respectively, spaced along the central $12.8$~m of the focal lines (see Fig.~\ref{fig:scimatic}).  The slightly different feed spacing for the three cylinders enhances the map-making performance by helping disentangle spurious grating lobes \cite{jzhang2016b}.

In the future, the Pathfinder cylinder instrument can be upgraded by simply adding more feed units and the associated electronics. The longer-term plan is to expand the Tianlai array to full scale once the principle of intensity-mapping is proven to work. The full scale Tianlai cylinder array will have a collecting area of $10^{4}\,\text{m}^{2}$, and $10^{3}$ receiver units. A forecast for its capability in measuring dark energy and constrain primordial non-Gaussianity can be found in \cite{Chen2012, Xu2015}.

\begin{figure}
\centering
\includegraphics[width=0.70\textwidth,trim =  0 10 10 10, clip]{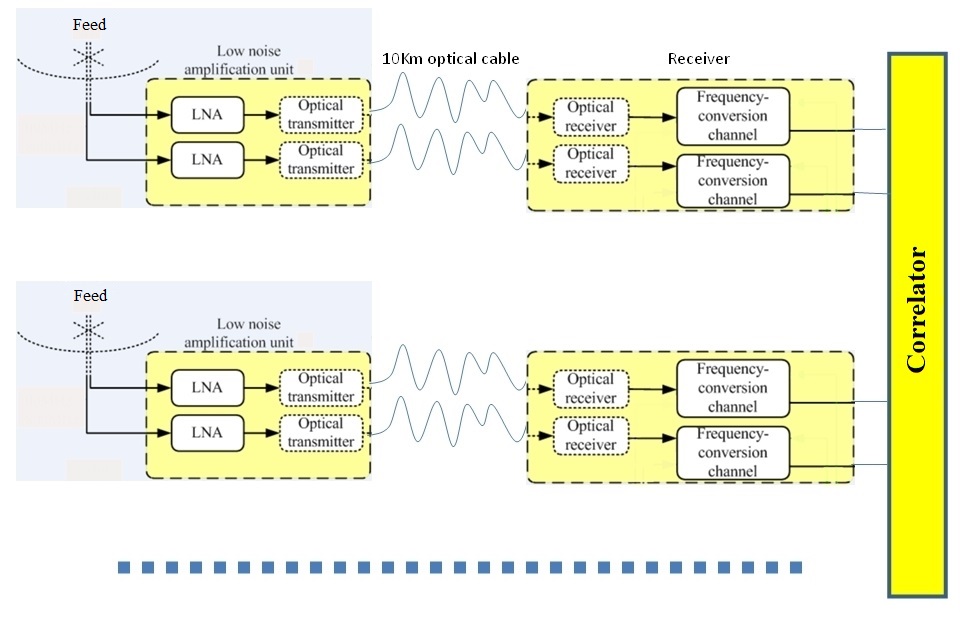}
\caption{\label{fig:receiver-chain} Schematic view of the analog signal path in the Tianlai Pathfinder experiment. Analog signals are transported over long optical fibers  from the antenna to electronic and correlator room, located in the base facility, more than 7~km away. (Figure adapted from Y. Wang et al.\cite{YWang2018}.) }
\end{figure}

\subsection{Electronic chain and correlator}
In Fig.~\ref{fig:receiver-chain} we show a schematic view of the overall architecture of the Tianlai electronic and data acquisition chain.  The analog signals from the feed, amplified by the LNA (Low Noise Amplifier) are converted 
to optical signals at the instrument site and sent to the base facility, more than 7~km away through optical fibers. In total, there are 192 analog signals from the cylinder array and 32 signals from the dish array. The optical signals are then converted back to electric signals, down-converted to the IF band [$135\sim235$~MHz] with a tunable local oscillator at 935~MHz, and fed to the correlator. 

The dish array signals are fed into an FPGA-based correlator, while the cylinder array signals are fed to a purpose-built Tianlai correlator, based on DSPs (Digital Signal Processors). Dish data represent 528 visibilities, for 512 frequency channels, $\sim 244 \, \mathrm{kHz}$ wide each, covering  $\sim 685 - 810 \, \mathrm{MHz}$ frequency band. The full data rate out of the dish array correlator represents $\sim 2 \,\mathrm{MB/s}$ with visibilities averaged over 1 second.

The Tianlai cylinder correlator is an FX-type correlator.  The 192 input signals are sampled at 250~MSPS, using 14 bit ADCs, with the 250~MHz clock synthesized from a GPS-stabilized 10~MHz clock. Input signals are distributed into groups of 8 signals, each group sampled by a mezzanine FMC board with 4 dual input ADC circuits. The sampled signals are then truncated to 8 bits and converted into frequency components through a 2048 point FFT implemented on Xilink Virtex-6 FPGA’s. Each sampling/FFT board has 2 Virtex-6 FPGA's, each one processing the data from an FMC ADC board.  The sampled data is then sent to correlator boards through data switch boards, each with 4 RapidIO switches, with transfer rates up to $ 6 \, \mathrm{Gbps/lane}$. Each correlator board has 8 Texas Instrument TMS320C6678 DSP and can perform up to $\sim 2.5 \times 10^9$ Multiply/Accumulate operations per second. The complex visibility data are then written as pairs of 32 bit IEEE floating point numbers to SATA disk arrays, with typical visibility averaging time of 4 second. 

The full cylinder correlator system has 12 sampling/FFT boards, 20 switch boards, 27 correlator boards and 1 control board. The total output data rate of the cylinder array correlator is 
$\sim 142 \, \mathrm{MB}$ per averaging time for 18,528 visibilities and 1,008 frequency channels, each 122~kHz wide ($\delta z=0.0002$), covering the frequency range from 686 to 809~MHz. 
   

\subsection{Noise source and calibration}

The Tianlai Pathfinder has a relatively low system sensitivity,
so the number of astronomical calibration sources available is limited and not regularly distributed on the sky. We set up an artificial noise source to monitor the variation of system gain and instrumental phase so that a relative calibration can be performed. 

A noise source diode is used to generate a broad band, Gaussian white noise spectrum, which is then amplified to achieve a high S/N ratio. A lightning protector and bandpass filter are inserted in the first stage after the antenna. The noise source is located at the top of a small hill about 120~m to the West of the cylinder array.  The emission pattern of the antenna (a discone antenna) for the noise source is omnidirectional.  The power of the noise diode and amplifiers are controlled by a power switch controlled by a TTL logic signal, sent periodically from the correlator.

We have taken special care to stabilize the temperature of the
noise source. The elements of the noise source system are all mounted on a copper board which has a dedicated thermostatic system that stabilizes the temperature to less than $0.2^\circ$C variations over a day to provide stable calibration of the system gain.

\begin{figure}
\includegraphics[width=0.49\textwidth,trim =  0 180 220 30, clip]{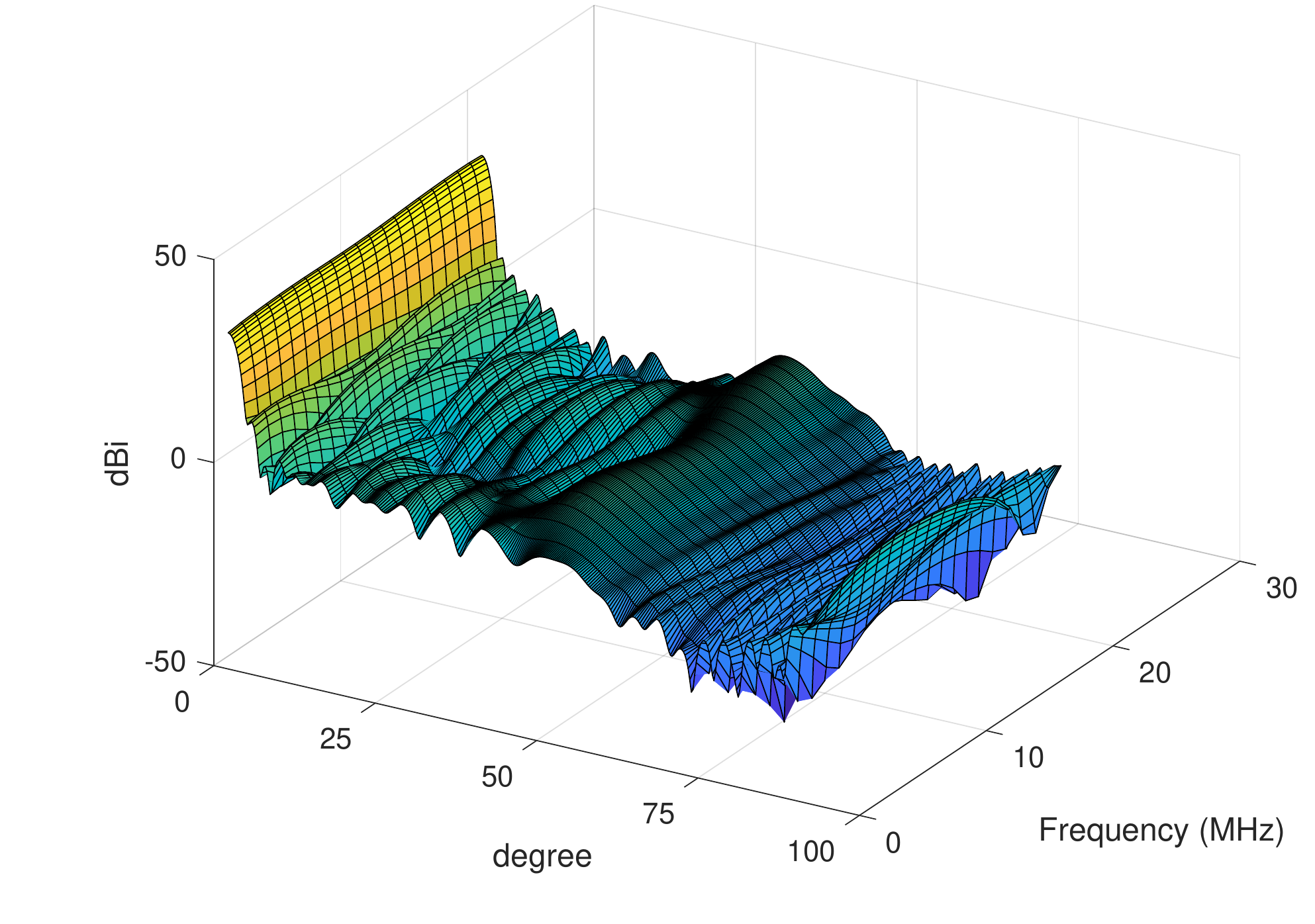}
\includegraphics[width=0.49\textwidth,trim =  0 50 0 80, clip]{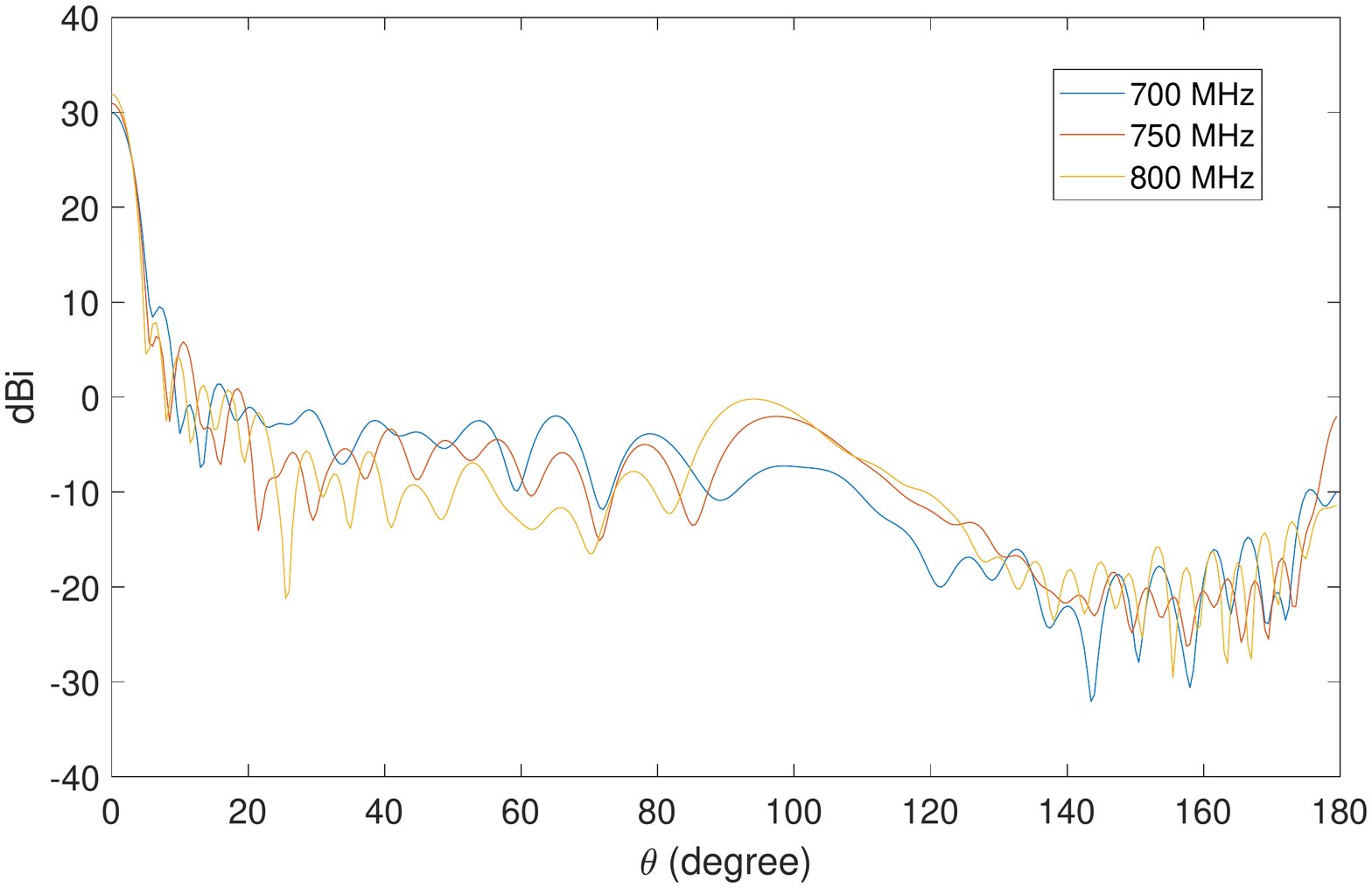}
\caption{Left: Simulated beam pattern for a Tianlai dish antenna as a function of both angle $\theta$ and frequency.   Right: Simulated beam pattern as a function of $\theta$ for 3 different frequencies, 700, 750 and 800~MHz. Each plot shows the absolute co-polar gain in dBi in the E-plane ($\phi =0$).  Angle $\theta$ is calculated from the center of the beam. 
The simulation shows that the antenna sidelobes vary significantly as a function of frequency.  
\label{fig:simulated_beam}}
\end{figure}

\section{Tianlai measurements and analysis}
\label{sec:description}

\subsection{Beam pattern analysis for the dish array }

Electromagnetic simulations for the scattering parameters and the beam pattern are performed using CST Microwave Studio, an electromagnetic simulation software package. We have used the CST transient solver and use a waveguide port on the $50~\Omega$ coaxial line leading to the excited dipole in the feed antenna. The accuracy of the solver was set to $-30$~dB, which means that the solver continued to calculate the field distribution and S-parameters until the electromagnetic field energy inside the structure had decayed to below $-30$~dB of the maximum energy inside the structure at any time.

We have calculated the beam pattern for different frequencies between $700$~MHz  to $800$~MHz in steps of $5$~MHz. Fig.~\ref{fig:simulated_beam} shows the  beam as a function of $\theta$ from the center of the beam for different frequency channels.  The plots show that the sidelobes of the beam vary significantly with frequency. 

\begin{figure}

\includegraphics[width=0.49\textwidth,trim =  0 50 0 80, clip]{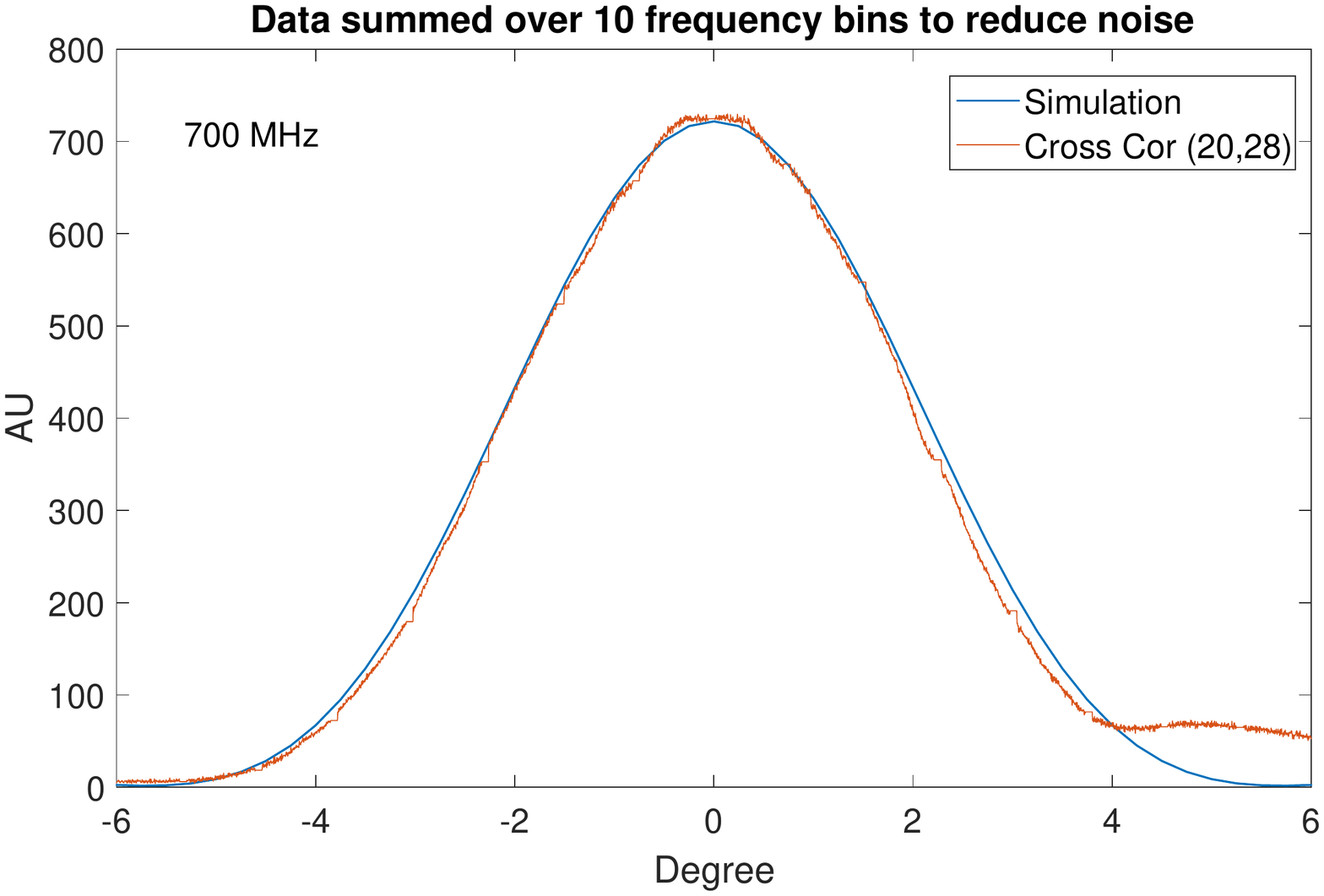}
\includegraphics[width=0.49\textwidth,trim =  0 50 0 80, clip]{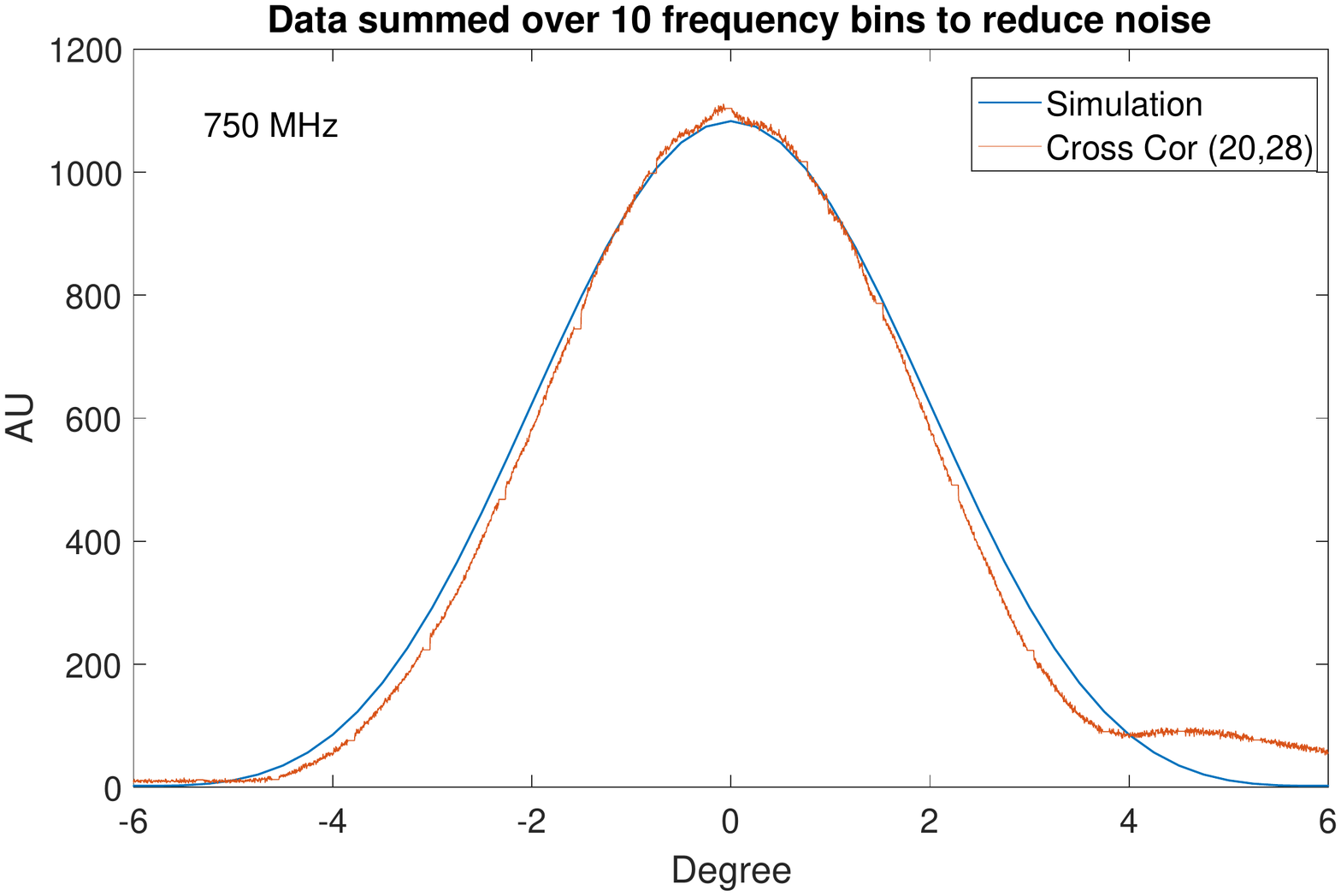}

\caption{\label{fig:cygnus_transit}
Plots comparing the simulated main beam pattern and the  pattern measured during the transit of Cygnus A.  We have matched the gain amplitudes by visual  inspection;  there are no other adjustable parameters. The units on the vertical axis are arbitrary. To reduce the noise we have averaged over 10  frequency bins around the given frequency. The agreement between the simulation and measurement is good. We have shown the comparison at two different frequencies- 700~MHz and 750~MHz. The `shoulder' on the RHS of the measured patterns is from diffuse emission from Cygnus X and the Galactic plane. (Here, `AU' stands for `Arbitrary Units.')}
\end{figure}

To test the accuracy of our beam simulations, we measure the patterns by fixing our telescope at a declination of $\sim 44.73^\circ$ and observing Cygnus A with the drift-scan technique.   Fig~\ref{fig:cygnus_transit} shows the simulated and measured patterns for frequency $700$~MHz and $750$~MHz. In each case, the amplitude of the simulated pattern is adjusted by eye to achieve a good fit and not by any parameter estimation method. The periodic stripes in the data come from the noise source, which is turned on at those times. We remove those points when plotting the data.  We have plotted the cross-correlation function because in the auto-correlation function there is another peak near Cygnus A, coming from distributed emission from Galactic plane in the vicinity of Cygnus X. Therefore, there is no single peak  with which we can match with the simulated beam. However, in the cross-correlation  function the distributed emission appears as a small hump just next to Cygnus A.  Therefore, it is possible to fit the simulation with the cross-correlation visibility.  We co-plot the visibility over the simulated beam pattern and the plots do match with a fairly good accuracy. 

\subsection{Cylinder antenna East-West Beam profile fitting}

We also simulate the beam patterns of the cylinder array; a detailed analysis appears in Cianciara {\it et al.} (2017) \cite{Cianciara2017}.  We have been able to measure the East-West beam profile of the cylinder arrays using the strong celestial point source Cygnus A. A careful comparison of the simulations with the measured patterns is underway.

If we take the source as an ideal point with flux $S_{c}$ at direction $\hat{n}_{0}$, then its observed visibility
(neglecting all other signals and noise) is 

\begin{equation}
V_{ij}^{0}=S_{c}\,g_{i}g_{j}^{*}A_{i}(\hat{n}_{0})A_{j}^{*}(\hat{n}_{0})e^{2\pi i\hat{n}_{0}\cdot(\vec{u}_{i}-\vec{u}_{j})}=S_{c}\,G_{i}G_{j}^{*}\label{eq:Vps}
\end{equation}
where $A_i(\hat{n})$ is the primary beam of feed i, $\vec{u}_i$ is the position vector of feed $i$ and $g_i$ is the (complex) gain factor of feed $i$.
\[
G_{i}=g_{i}A_{i}(\hat{n}_{0})e^{2\pi i\hat{n}_{0}.\vec{u}_{i}}\;.
\]

\noindent Writing Eq.$~(\ref{eq:Vps})$ in matrix form, we get
\[
V_{0}=S_{c}\,\vec{G}\vec{G}^{\dagger}\;.
\]

\noindent The vector $\vec{G}$ can be obtained by an eigen-analysis of $V_{0}$. 

After solving for $\vec{G}$, we have 
\[
|G_{i}|=|g_{i}|A_{i}(\hat{n}_{0})\propto A_{i}(\hat{n}_{0})
\]
 if the amplitude of the gain $g_{i}$ is stable. This gives an East-West
beam profile along the transit track of the point source.

\begin{figure}
\includegraphics[width=0.49\textwidth,trim =  0 0 40 40, clip]{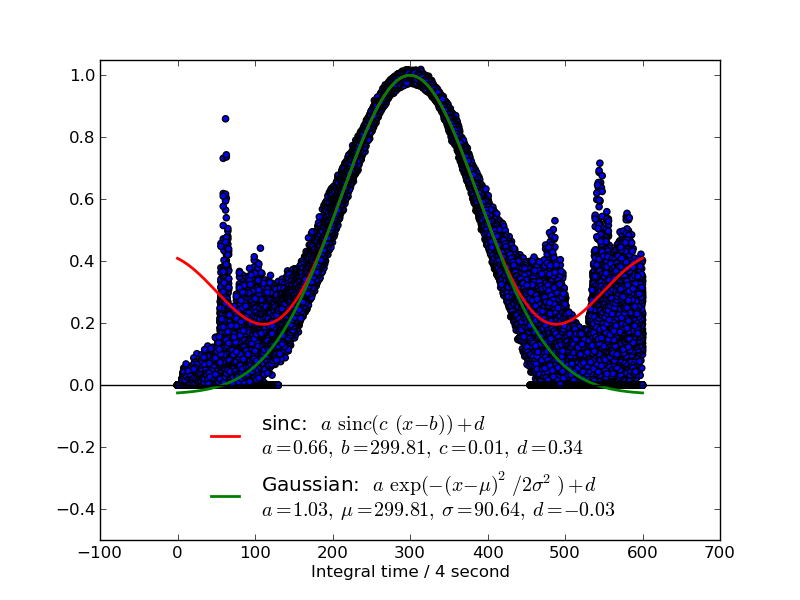}
\includegraphics[width=0.49\textwidth,trim =  0 0 40 40, clip]{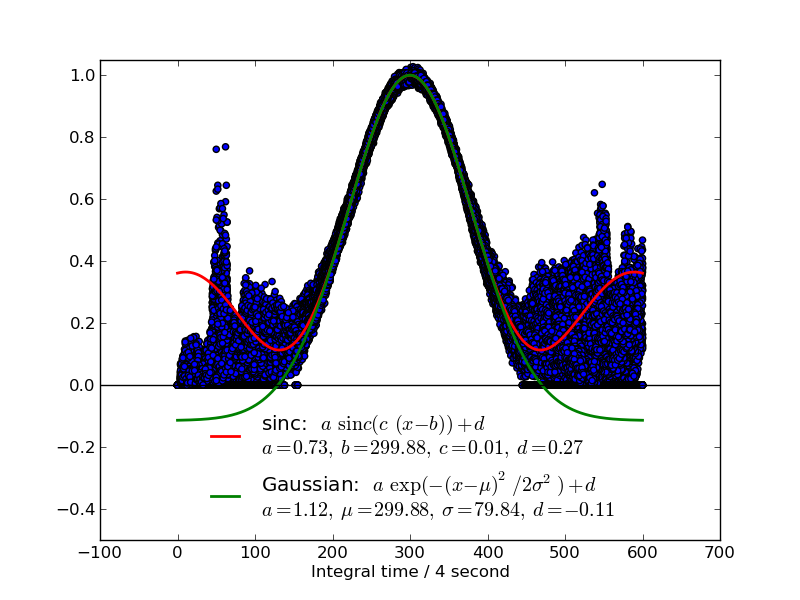}
\caption{\label{fig:Bf} We used Cygnus A as a calibration source to measure the East-West beam profile of the cylinder array for all 96 feeds. The analysis is done at 
750~MHz. Each of the feeds has a dual polarization receiver. Left: Analysis for $XX$ polarization (transverse to the cylinder axis), 
Right: Analysis for $YY$ polarization (parallel to the cylinder axis). In each case, the centers of the beam profiles are co-aligned.}
\end{figure}

We calculate the East-West beam profile for the feeds installed in the cylinder array.  There are a total of 96 dual polarization receivers.  We calculate both the $XX$  (transverse to the cylinder axis) and $YY$ (oriented parallel to the cylinder axis) polarizations of all 96 feeds. Then we match the peaks of all the beams.  Our result is shown in Fig.~\ref{fig:Bf}. The analysis is done at $750$~MHz.
We have also calculated the FWHM of the beam main lobe. Our results show that the FWHM is about $3.6^\circ$ for the $XX$ polarization, and about $3.15^\circ$ for the $YY$ polarization.

\begin{figure}[h]
\includegraphics[width=0.98\textwidth,trim =  0 0 0 0, clip]{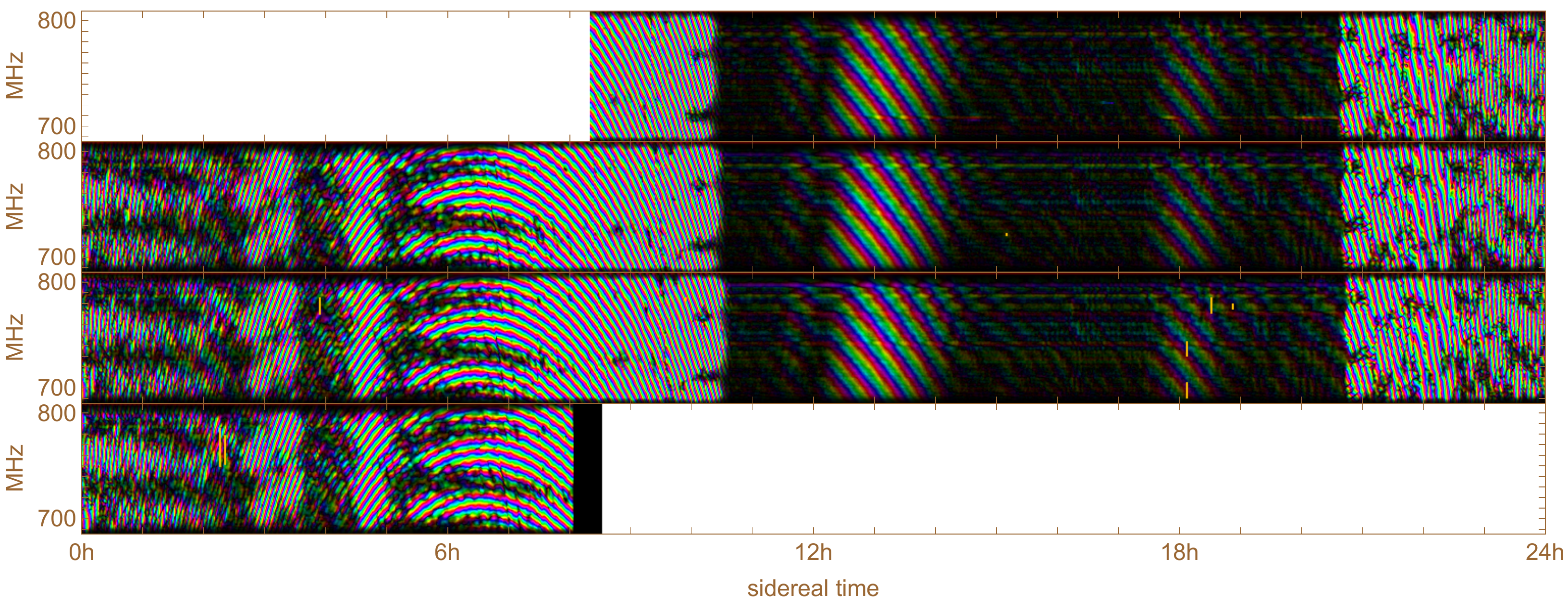}
\caption{\label{fig:StebbinsPlot} Raw visibilities are plotted as a function of sidereal time and frequency from one of 528 baselines of the dish array while pointed near to the North Celestial Pole. Visibilities are shown for four consecutive days. These are uncalibrated visibilities with no RFI subtraction and averaged over 1 minute and 1~MHz.  In this image intensity gives the amplitude of the visibility and hue gives the complex phase (the fringe pattern).  The time variation is primarily due to Earth rotation and the visibilities repeat every sidereal day with high accuracy. The day-to-day repeatability of the visibility shows the stability of the system. The period of the fringes indicates the declination of the sources: low declination sources move rapidly on the sky whereas high declination sources move more slowly. The Sun is above the horizon from 10.5~hr to 20.5~hr and creates the rapidly-varying bright fringes dominating the left and right sides of the image.  More slowly-varying and dimmer fringes appear when the Sun has set.  These nighttime fringes are from sources at high declination and closer to the beam center.  The beam pattern near the beam center is smooth and produces a relatively clean nighttime fringe pattern whereas the beam pattern toward the Sun (far off axis) is very complicated and results in a daytime fringe pattern with complicated gain modulation in both frequency and time which appear as bright and dark patches. The horizontal striping evident during the night is a result of RFI.  The occasional orange spots are from correlator glitches.}
\end{figure}

\subsection{A quick look at visibility data}

In Fig.~\ref{fig:StebbinsPlot} we have plotted the raw visibility for one baseline of the dish array over several days.  The antennas were pointed at the North Celestial Pole (NCP) and the Sun is approximately $70^\circ$ from the beam center.  During daytime the fringes mainly come from the Sun in the far side lobes of the antenna pattern.  Removal of the solar signal at the required level seems difficult so we anticipate obtaining 21~cm survey data at night and using the daytime data to study the beam pattern. The Sun moves through 48 degrees of declination over the course of a year so this data could be used to precisely measure the off-axis beam pattern across a wide range of polar angles.  The fringes during the nighttime come from several bright sources near the NCP.



\begin{figure}[htbp]
\centering
\includegraphics[width=0.75\textwidth]{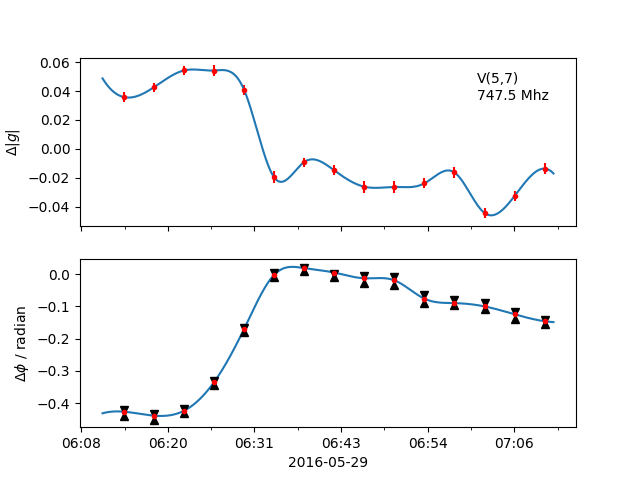}
\caption{\label{fig:gainChange} Complex gain amplitude (top) and phase (bottom) variations over time, as determined from the noise source signal, for a particular visibility at 747.5 MHz. Top: fractional gain, 
Bottom: complex gain phase $\Delta \Phi$. 
Triangles on the phase plot	show the  
phases of the adjacent frequency bins 
(one higher and one lower). 
}
\end{figure}

 In Fig.~\ref{fig:gainChange} we have plotted the gain and the phase change over an hour for a particular baseline, using a noise source signal exciting the feed through sidelobes.
We observe small variations in the gain amplitude and phase. The magnitude and the structure of the gain and phase variations are still not well understood. We see similar structures in the gain amplitude variations for different baselines, suggesting that the amplitude variations may be due to the noise source itself. However, the variation in phase is still unexplained. Since the time that this data was taken, the noise source has been thermally stabilised.  

\section{Tianlai data-processing pipeline: \texttt{tlpipe}}
\label{sec:tlpipe}

\begin{figure}[h]
\includegraphics[width=0.98\textwidth]{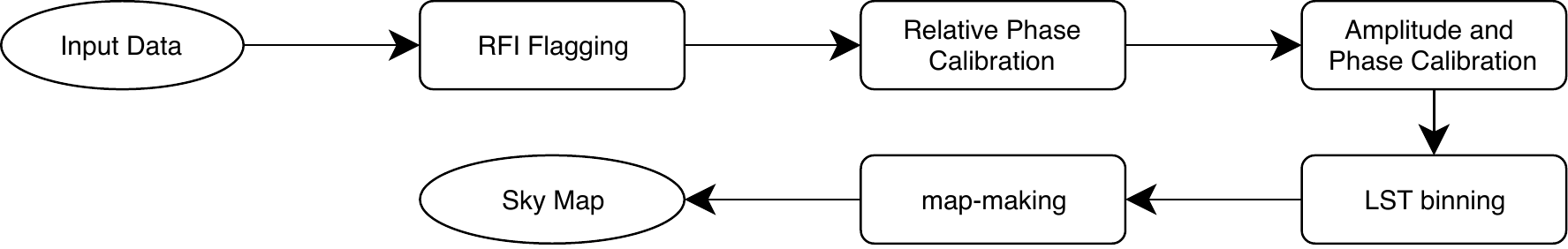}
\caption{\label{fig:flow} Schematic of the data processing pipeline implemented in the \texttt{tlpipe} package.} 
\end{figure}

The Tianlai data processing pipeline, \verb!tlpipe!\footnote{\url{https://github.com/TianlaiProject/tlpipe}}, is a Python package specifically developed for the Tianlai array. The pipeline is almost complete for the early stage of data processing tasks, from reading data from raw observing data files, to the final map-making. The basic data processing procedures like RFI flagging, noise source calibration, point source calibration, data binning, and map-making are implemented, along with various auxiliary analysis
utilities like data selection, transformation, visualization and so on. A schematic of the data processing pipeline implemented in the package is shown in Fig.~\ref{fig:flow}. 

The raw observational data are stored in HDF5 file format, which supports parallel I/O operations. To maintain consistency of the data file format, the pipeline uses the HDF5 file format for all data input from disks and writes the processed data and all intermediate data to HDF5 format.  To support the manipulation of large amounts of data and improve data processing efficiency, \texttt{tlpipe} is developed based on the MPI framework,  but the package also works in a non-MPI environment. Our tests show that it works well under a broad range of scales from a single process to about $10^{4}$ MPI processes.  The package is built on the Python scientific computing stack including numpy, scipy, matplotlib, h5py, etc., for highly effective computing.  Some performance-critical parts that can not be well implemented by using the numpy array operations are statically compiled using the Cython package.

The package consists of three main components that interact with each other: the task executing framework, the tasks, and the data container.  The task executing framework, also called the task manager, controls the execution of the tasks. A task is usually an individual or independent data processing step, which is implemented according to a few common interfaces to make it executable by the task manager. The data container holds the data and some descriptive metadata to be processed by a task. A data processing pipeline usually consists of several tasks that will be executed in a specified order (some of the tasks may be iteratively executed many times according to the data processing requirement).  Setting-up, executing the tasks, and executing the flow control are fulfilled by the task manager according to the settings in an input parameter file provided by the user. The parameters set in the input parameter file fully determine the pipeline control flow and the behavior of each task in the pipeline. 
Full data processing can be submitted and run on a personal computer, a cluster, or a supercomputer with only one parameter file. No other intermittent human interaction  is required until it is done or it crashes due to some error.

The package is developed using the git version control tool. It is an open-source project and can be downloaded from the GitHub website. 

\section{Discussion and Future Plans}
\label{sec:conclusion}

The Tianlai Pathfinder is currently being commissioned. In this paper, we have discussed the present status of the instrument, some of the issues that we have found in the current data, and our attempts to resolve them.  Before undertaking scientific surveys, we plan to acquire more data of known sources to calibrate the instrument. 

Another future goal is to study the beam using a drone outfitted with a radio source. In order to achieve the required accuracy for $21~$cm intensity-mapping, one needs to understand and calibrate the beam pattern of each telescope very well.  
Small deviations of the mechanical configuration or the environment ({\it e.g.} temperature) of the telescopes can change the beam pattern and introduce systematic errors in the astronomical measurement.  The drone study can offer high measurement accuracy at a relatively  moderate cost. We have purchased a drone ({\it DJI Matrice-600 Pro}),  which is capable of carrying a payload mass as large as $6$~kg. It can reach an altitude of $2500~$m with positional accuracy of $\sim 1~$cm. We hope to perform the first beam-mapping tests with the drone at Tianlai in the summer of 2018.

Tianlai's pointable dish array allows us to point the dishes toward the North Celestial Pole (NCP) where the total area surveyed by a transit telescope over a day is smallest.  Here Tianlai can integrate down to a low noise temperature map in the shortest amount of observation time. We expect to produce a high signal-to-noise map of the 21~cm emission in this region.  To aid in calibrating the dish array for accurate map making near the NCP we expect to obtain supplemental in-band calibrations of all the bright radio point sources near the NCP. 

We are currently observing the NCP with the $700-800\,$MHz band for 21~cm redshifts $z\sim1$ and later we plan to retune to a band with redshift $z\sim0.15$ and repeat these observations.  The lower redshift 21~cm survey will map the same volume as that occupied by the galaxies in the Northern Celestial Cap optical photometric survey \cite{Gorbikov2014}.  We can thus compare the 3D 21~cm map to a 3D galaxy map constructed from either photometric redshifts or preferably from optical spectroscopic  redshifts we expect to obtain with supplemental observations.  An accurate reconstruction of the 3D distribution of optical galaxies with the 21~cm map obtained by intensity mapping will validate the efficacy of this method.  

Finally, in addition to redshifted $21~\text{cm}$ intensity-mapping observations, our surveys may also be used for other observations, such as searching for $21\,$cm absorbers, fast radio bursts (FRB), and electromagnetic counterparts of gravitational wave events. 

\section{Acknowledgements}
Work at UW-Madison and Fermilab is partially supported by
NSF Award AST-1616554. Work at NAOC is supported by the MOST grants 2016YFE0100300 and 2012AA121701, the NSFC 11473044 and 11633004, the CAS QYZDJ-SSW-SLH017. Part of the Computation is performed on the Tianhe-2 supercomputer with the support of NSFC grant U1501501.
Authors affiliated with French institutions acknowledge partial support from CNRS (IN2P3 \& INSU), Observatoire de Paris and from Irfu/CEA.  This document was prepared by Tianlai Collaboration using the resources of the Fermi National Accelerator Laboratory (Fermilab), a U.S. Department of Energy, Office of Science, HEP User Facility. Fermilab is managed by Fermi Research Alliance, LLC (FRA), acting under Contract No. DE-AC02-07CH11359.

\bibliographystyle{plain}
\bibliography{bibliography}
\end{document}